\documentclass[useAMS,twocolumn]{mn2e}
\usepackage{amssymb}
\usepackage{graphicx}

\usepackage[dvips]{color}

\title[Does the circularization radius exist or not?]
 {Does the circularization radius exist or not for low angular momentum accretion? }
\author[Bu \& Yuan]
{De-Fu Bu$^{1}$\thanks{E-mail:dfbu@shao.ac.cn} and Feng
Yuan$^{1}$\thanks{E-mail:fyuan@shao.ac.cn}\\
$^{1}$Key Laboratory for Research in Galaxies and Cosmology,
Shanghai Astronomical Observatory,\\ Chinese Academy of Sciences, 80
Nandan Road, Shanghai 200030, China\\}

\date{Accepted . Received ; in original form}

\begin{document}

\maketitle

\begin{abstract}
If the specific angular momentum of accretion gas at large radius is
small compared to the local Keplerian value, one usually believes
that there exists a ``circularization  radius'' beyond which the
angular momentum of accretion flow is almost a constant while within
which a disk is formed and the angular momentum roughly follows the
Keplerian distribution. In this paper, we perform numerical
simulations to study whether the picture above is correct in the
context of hot accretion flow. We find that for a steady accretion
flow, the ``circularization  radius'' does not exist and the angular
momentum profile will be smooth throughout the flow. However, for
transient accretion systems, such as the tidal disruption of a star
by a black hole, a ``turning point'' should exist in the radial
profile of the angular momentum, which is conceptually similar to
the ``circularization radius''. At this radius, the viscous
timescale equals the life time of the accretion event. The specific
angular momentum is close to Keplerian within this radius,  while
beyond this radius the angular momentum is roughly constant.
\end{abstract}


\section{INTRODUCTION}
Accretion is an important physical process in astrophysics.
According to the temperature of the accretion flow, accretion disk
models can be broadly divided into two series, namely cold and hot.
The representative one of the former is the standard thin disk
(Shakura \& Sunyaev 1973). In this model, the gas can cool very
efficiently, the temperature is much lower than the virial
temperature so the disk is geometrically thin and the radiative
efficiency is high. The second one is hot accretion flow, such as
advection-dominated accretion flows (ADAFs; Narayan \& Yi 1994;
1995; Abramowicz et al. 1995; see Yuan \& Narayan 2014 for the
recent review of hot accretion flow and its astrophysical
applications). Different from the thin disk,  the temperature is
almost virial, the radiative efficiency of hot accretion flow
increases with accretion rate (Xie \& Yuan 2012). The most important
progress in the field of hot accretion flow in recent years is
perhaps the finding of the strong outflow or wind which is launched
at any radius throughout the disk (Yuan, Bu \& Wu 2012; Narayan et
al. 2012; Li, Ostriker \& Sunyaev 2013). Recently, this theoretical
prediction has been confirmed by the 3 million seconds {\it Chandra}
observation to the supermassive black hole in our Galactic center,
Sgr A* (Wang et al. 2013). This result is interesting because wind
is not only an important factor in accretion physics but also plays
an important role in AGN feedback (e.g., Ostriker et al. 2010; Gan
et al. 2014).

In this paper, we address the question of low-angular momentum
accretion. Such kind of accretion is common in the universe. For
example, many elliptical galaxies, including M87, have stellar
populations with small average spin that is insufficient to create
Keplerian disks near the Bondi radius (see, e.g., Inogamov \&
Sunyaev 2010). If the specific angular momentum of the accretion
flow at large radius is very low compared to the local Keplerian
value, a popular picture people have is as follows. Denoting the
specific angular momentum of the accretion flow as $l_0$. The flow
will keep their angular momentum $l_0$ from the outer boundary until
a ``circularization  radius'' $R_{\rm cir}$ determined by $l_{\rm
k}(R_{\rm cir})=l_0$, here $l_{\rm k}$ is the specific Keplerian
angular momentum. Within $R_{\rm cir}$, the angular momentum of
accretion flow will roughly follow the Keplerian distribution.

The concept of circularization radius was perhaps first proposed in
the context of accretion disks in binary systems as the
characteristic radius at which the mass transfer stream from the
companion will orbit at first after interacting with either the
stream itself or with the disk if one was already present. Later
this concept was extended and often applied now to the steady
accretion flow in general in the literature (e.g., Melia, Liu \&
Coker 2001). We will show in this paper that the `circularization
radius'' does not exist for steady accretion. Our idea is stimulated
by Yuan (1999). In that work, we calculated the one-dimensional
steady global solution of hot accretion flow with various outer
boundary conditions. Specifically, when the specific angular
momentum is low, it was found that the radial profile of the angular
momentum is quite smooth throughout the accretion flow and there is
no ``circularization radius''. The accretion flow simply spirals in
and a Keplerian disk is never formed. Such kind of accretion is
called Bondi-like accretion since the sonic radius is usually quite
large (see also Abramowicz \& Zurek 1981 for the study of Bondi-like
accretion for an inviscid flow).

In this paper, we will first analyze the physical reason for the
absence of the circularization radius (\S2). Then we perform
hydrodynamic (HD) and magnetohydrodynamic (MHD) numerical
simulations to investigate this problem (\S3). We do find the
existence of a ``turning point'' in the angular momentum
distribution at the begin of evolution, which is similar to the
``circularization radius''. However, the ``turning point'' moves
outward with time in the viscous timescale. Thus, provide that there
is long enough time for the accretion flows to evolve and a steady
state is reached, the angular momentum will be smooth throughout the
accretion flow. We summarize our results in \S 4.

\section{Analytical considerations}

It is now widely accepted that the mechanism of angular momentum
transport in ionized accretion flows is the magnetorotational
instability (MRI; Balbus \& Hawley 1991; 1998). In the hydrodynamic
studies, we usualy use a Newtonian stress tensor $\mathbf{T}$ to
mimic this turbulent stress. Following Stone, Pringle \& Begleman
(1999, hereafter SPB99), we assume that the only non-zero components
of $\mathbf {T}$ are the azimuthal components,
\begin{equation}
  T_{r\phi} = \mu r \frac{\partial}{\partial r}
    \left( \frac{v_{\phi}}{r} \right),
\end{equation}
\begin{equation}
  T_{\theta\phi} = \frac{\mu \sin \theta}{r} \frac{\partial}{\partial
  \theta} \left( \frac{v_{\phi}}{\sin \theta} \right) .
\end{equation}
This is because the MRI is driven only by the shear associated with
orbital dynamics. Other components of the stress are much smaller
than the azimuthal components (Stone \& Pringle 2001). We adopt the
coefficient of shear viscosity $\mu=\nu\rho$. For one-dimensional
case, we focus only on $T_{r\phi}$ since only this component
appears in the radial component of angular momentum equation. We
assume $\nu=\alpha (r/r_s)^{1/2}(2GM/c)$, where $M$ is the center
black hole mass, $G$ is the gravitational constant and $r_s \equiv
2GM/c^2$ is the gravitational radius. For a ``normal'' accretion
disk, the angular velocity $\Omega \propto r^{-3/2}$. Based on our
assumption, $T_{r\phi} \propto \rho r^{1/2} \Omega_k (r/\Omega_k
d\Omega/dr) \propto \rho r^{-1}$. For a hot accretion flow, the gas
temerature is virial; so the square of sound speed $c_s^2 \propto
r^{-1}$. Therefore, the viscous tensor used in this paper $T_{r\phi}
\propto \rho c_s^2 \propto \alpha p$, which is the usual
``$\alpha$''-viscosity description. We choose this type of stress description because
it is in good consistency with the MHD numerical simulation results, namely
the magnetic stress is proportional to the total pressure (e.g.,
Hirose, Krolik \& Blaes 2009). Therefore, the viscosity description adopted above can well represent the real case.

Based on the above knowledge, we now analytically discuss whether the ``circularization radius'' scenario is correct or not. For a
steady state, the radial angular momentum transfer equation can be
simply written as
\begin{equation}
\rho v_r \frac{\partial l}{\partial
r}=\frac{1}{r}\frac{\partial}{\partial r}(r^2 T_{r\phi}), \label{angulareq}
\end{equation}
where $l$ is specific angular momentum. In the ``circularization radius'' scenario, the specific angular momentum at $R\ga R_{\rm cir}$ is constant. Therefore, for this region the
left hand side of eq. (\ref{angulareq}) is zero. In order to satisfy eq. (\ref{angulareq}), the
right hand side should also be zero. This requires $r^2 T_{r\phi}$
is constant of radius. Therefore, $T_{r\phi} \propto r^{-2}$. From
equation (1), if the angular momentum of the disk is a constant of
radius, $T_{r\phi} \propto \rho * r^{-3/2}$. If equation (3) is
satisfied, the density profile should be $\rho \propto r^{-1/2}$. In
the region $r\geq R_c$, the angular momentum is negligible compared
to the Keplerian value. We have done test and find that if the
angular momentum of an hydrodynamic flow is negligible, the density
profile is $\rho \propto r^{-(1.3-1.5)}$ (Bu et al. 2013).
Therefore, equation (3) can not be satisfied and thus the ``circularization
radius'' scenario is problematic.

\section{Numerical simulations}

\subsection{Equations}
In this section, we further carry out both HD and MHD simulations using the
ZEUS-2D code (Stone \& Norman 1992a,1992b) to study the angular
momentum profile of small angular momentum gas accretion. The
equations are exactly the same as those of SPB99 and Stone \&
Pringle (2001). For the convenience of readers, we copy them here.
The HD equations are
\begin{equation}
\frac{d\rho}{dt}+\rho\nabla\cdot \mathbf{v}=0,\label{cont}
\end{equation}
\begin{equation}
\rho\frac{d\mathbf{v}}{dt}=-\nabla p-\rho\nabla
\psi+\nabla\cdot\mathbf{T}, \label{rmon}
\end{equation}
\begin{equation}
\rho\frac{d(e/\rho)}{dt}=-p\nabla\cdot\mathbf{v}+\mathbf{T}^2/\mu.
\label{rmon}
\end{equation}
The MHD equations are
\begin{equation}
\frac{d\rho}{dt}+\rho\nabla\cdot \mathbf{v}=0,\label{cont}
\end{equation}
\begin{equation}
\rho\frac{d\mathbf{v}}{dt}=-\nabla p-\rho\nabla
\psi+\frac{1}{4\pi}(\nabla \times \mathbf {B}) \times \mathbf {B},
\label{rmon}
\end{equation}
\begin{equation}
\rho\frac{d(e/\rho)}{dt}=-p\nabla\cdot\mathbf{v}+\eta \mathbf {J}^2,
\label{rmon}
\end{equation}
\begin{equation}
\frac{\partial \mathbf {B}}{\partial t}=\nabla \times (\mathbf {v}
\times \mathbf {B}-\eta \mathbf {J}).
\end{equation}

In the above equations, $\rho$, $p$, $\mathbf{v}$, $\psi$, $e$,
$\mathbf{B}$ and $\mathbf{J}(=(c/4\pi)\nabla \times \mathbf {B})$ are
density, pressure, velocity, gravitational potential, internal
energy, magnetic field and the current density, respectively. The
viscous stress tensor in equations (5) and (6) is expressed in
equations (1) and (2). $d/dt(\equiv
\partial / \partial t+ \mathbf{v} \cdot \nabla)$ denotes the Lagrangian
time derivative. We adopt an equation of state of ideal gas
$p=(\gamma -1)e$, and set $\gamma =5/3$. We use spherical
coordinates $(r, \theta, \phi)$ to solve the equations above.

We use the pseudo-Newtonian potential to mimic the general relativistic
effects, $\psi=-GM/(r-r_s)$. The self gravity of the gas is
neglected. In this paper, we set $GM=r_s=1$. We use the gravitational radius to
normalize the length scale. Time is in unit of the Keplerian orbital
time at $100r_s$.

In the MHD equations, the final terms in equations (9) and (10) are
the magnetic heating and dissipation rate mediated by a finite
resistivity $\eta$. Since the energy equation here is actually an
internal energy equation, numerical reconnection inevitably results
in loss of energy from the system. By adding the anomalous
resistivity $\eta$, the energy loss can be captured in the form of
heating in the current sheet (Stone \& Pringle 2001). The exact form
of $\eta$ is same as that used by Stone \& Pringle (2001).

\subsection{Model setup}

We carry out two models, models A (for HD) and B (for MHD). In model
A, we initialize our simulation as follows. The radial velocity,
internal energy and density of the accretion flow are adopted from
the hydrodynamic Bondi solution. We set the $\theta$ component of
the velocity $v_\theta=0$. For the rotation velocity, when $r>20
r_s$ and $45^{\circ} < \theta < 135^{\circ}$, we set $v_\phi =
l_0/r\sin{\theta}$ where $l_0$ is a constant. Within $20r_s$, the
angular momentum is zero. The initial distribution of the specific
angular momentum at $r>20r_s$ is shown by the dotted line in Fig. 1.
When we solve the Bondi solution, we set $\rho_\infty=1$ and the
ratio between the sound speed at infinity to the light speed is that
$2c_\infty^2/c^2=10^{-3}$. According to our settings, the ratio
between Bondi radius and gravitational radius $r_B/r_s=1000$. Our
purpose of setting such a large $c_\infty$ is that the Bondi radius
is small and is within our simulation domain. In this case, we can
study the accretion flow from close to the black hole to the Bondi
radius. In a realistic system, the Bondi radius should be much
larger than that used in this paper. We set $\alpha=0.05$ in model
A.

The initial conditions for the MHD model B are as follows. The
initial hydrodynamic variables of density, velocities and gas
internal energy are same as those in model A. The initial magnetic
field is generated by using a vector potential $\mathbf {B}=\nabla
\times \mathbf {A}$. The configuration of the magnetic field is same
as that used in Proga \& Begelman (2003): a purely radial field
defined by the potential $\mathbf{A}=(A_r=0, A_\theta=0, A_\phi=A_0
\cos\theta/r \sin\theta)$. We scale the magnitude of the magnetic
field by a parameter $\beta_0=8\pi p(r_0)/ \mathbf{B}^2$, with
$p(r_0)$ being the gas pressure at the outer boundary. In this case,
$A_0=sign(\cos \theta)\sqrt{(8\pi p(r_0)/\beta_0)}r_0^2$. In this
paper, we set $\beta_0=10^6$. The magnetic field strength is minimum
at the outer boundary and increases inward.

Our inner and outer radial boundaries are located at $1.5r_s$ and
$1200r_s$, respectively. The polar range is $0 \leq \theta \leq
\pi$. We divided the computational domain into $144 \times 100$
grids. We adopt non-uniform grid in the radial direction
$(\bigtriangleup r)_{i+1} / (\bigtriangleup r)_{i} = 1.05$. A
uniform polar grid extends from $\theta=0$ to $\pi$. At the poles,
we use axisymmetric boundary conditions. At the inner radial
boundary, we use outflow boundary conditions. At the radial outer
boundary, for hydrodynamic variables, we use inflow/outflow boundary
conditions (just copy the last active zone variables to ghost zone).
In the MHD model B, we fix the poloidal magnetic field to be its
initial configuration at the last zone in the radial direction.
While, for the toroidal magnetic field, we allow it to float.

\subsection{Results}

\begin{figure}
\includegraphics[width=9.0cm]{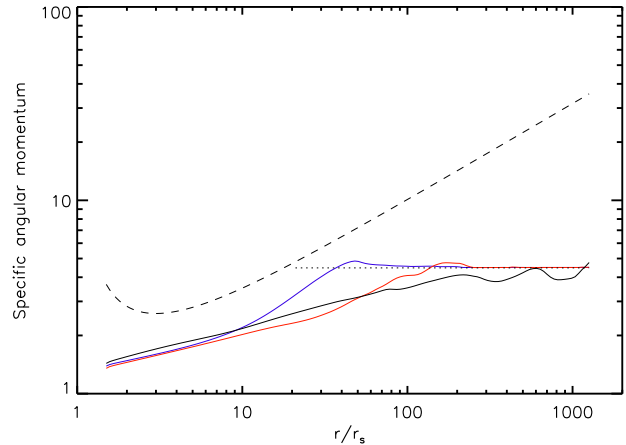}
\caption{The time evolution of the radial profile of the specific
angular momentum of accretion flow at the equatorial plane (averaged
from $\theta=84^{\circ}-96^{\circ}$) for model A. The dashed line
shows the Keplerian angular momentum. The dotted line shows the
initial angular momentum profile. The blue, red, and black lines
show the  angular momentum profiles at $t=0.03, 0.3$ and 22,
respectively. }
\end{figure}

\begin{figure}
\includegraphics[width=9.0cm]{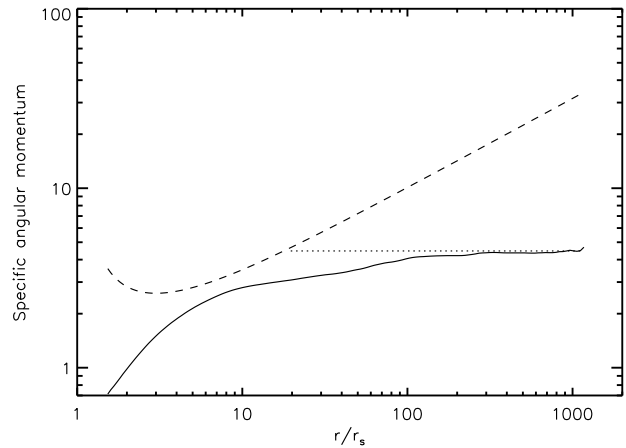}
\caption{The radial profile of the specific angular momentum of the
accretion flow at the equatorial plane (averaged from
$\theta=84^{\circ}-96^{\circ}$) for model B when a quasi-steady
state is reached. Because the fluctuation in this model is much
larger than that in model A, we use 50 data files cover 3 orbital
times (from $t=13$ to 16) to average the data when drawing this
figure. The dashed line shows the Keplerian angular momentum. The
dotted line shows the initial angular momentum profile.}
\end{figure}

In our models, we set $l_0=l_k (20r_s)$, here $l_k$ is the specific
angular momentum. This means that if the ``circularization radius''
scenario is correct, we should expect that the accretion flow
remains a constant angular momentum ($l_0$) outside of $20r_s$, they
forms  a Keplerian disk at  $\sim 20r_s$. We now check whether this
picture is true using simulations.

Fig. 1 shows the evolution of the radial profile of angular momentum
for model A which uses the usual ``$\alpha$''-viscosity description.
The dotted, blue, red, and black solid lines show the profiles at
time $t=0, 0.03, 0.3$ and $22$, respectively. At $t=22$, the
accretion flow inside of $100r_s$ has achieved a steady state. From
this figure, we can see that at the beginning of simulation,
$t=0.03$, the profile of angular momentum does look like to be
consistent with the ``circularization'' scenario, a ``turning
point'' is evident in the blue solid line which reminds us the
presence of the ``circularization radius''. However, the profile
quickly evolves with time. From the movie of evolution of accretion
flow we make, we find that the ``turning point'' in the profile
moves outward. Finally, when a steady state is reached, the profile
is quite smooth throughout the whole region and the ``turning
point'' completely disappears. The angular momentum at the outer
region $r> 1000r_s$ is larger than the initial value, this is
because of the angular momentum transfer from small to large radii.
Define the viscous time scale as $\tau_v=r^2/\nu$, we find that the
time needed to achieve a state that the angular momentum is smooth
throughout the flow is approximately $1/3$ times the viscous
timescale at the Bondi radius.

We have tested whether the value of $\alpha$ and the r-dependence of
viscosity affect our result by performing two test simulations. In
the first test, we assume the r-dependence of viscosity is same as
that in model A, but the value of $\alpha$ is five times smaller
than that in model A.  In the second test, we assume $\alpha$ is
same as that in model A, but $\nu \propto r$. We find that the
angular momentum profile in the two tests is very similar to that in
model A when a steady state is achieved.

For model B, we have checked that in the quasi-steady state, our
resolution in the inner region can resolve the fastest growth
wavelength of MRI for the actual angular velocity profile and
magnetic field strength. Fig. 2 shows the angular momentum profile
for model B in the quasi-steady state. The evolution of the profile
is similar to that shown in Fig. 1 and is not shown here. The
angular momentum profile is very smooth in the region $r<300r_s$.
The ``turning point'' moved to 300$r_s$ at the end of the
simulation. It indicates that our simulation has not run enough time
to reach a steady state at all radii. We expect that if we run the
simulation for longer time, the ``turning point'' will move outward
further until the angular momentum profile is smooth throughout the
flow. In the region $r < 6r_s$, the angular momentum decreases
quickly inward. The reason is as follows. Inside $6r_s$, the
magnetic field is very strong because of the accumulation of
magnetic flux in the simulation (the magnetic pressure is actually
higher than the gas pressure). Strong magnetic field results in
strong Maxwell stress in this region which can transport the angular
momentum very efficiently and makes the angular momentum inside
$6r_s$ decreasing quickly inward.

We note that Proga \& Begelman (2003) also studied the small angular
momentum gas accretion. Although the profile of angular momentum is
not explicitly shown in their paper, they do show that the
rotational velocity profile. From their Fig. 9, it seems that there
exists a ``circularization radius''. We suspect that this is because
they have not run the simulation for long enough time.

\section{SUMMARY}
A popular picture people have for the accretion of low-angular
momentum flow is that a ``circularization'' radius should exist.
Outside of this radius, the angular momentum of accretion flow is
constant while inside of this radius the angular momentum roughly
follow the Keplerian value thus a disk is formed. In this  paper, by
performing numerical simulations, we show that this picture is not
correct. We find that for low angular momentum gas accretion,
initially, there is a ``turning point'' at the ``circularization''
radius in the angular momentum distribution. But this ``turning
point'' will move outward with time. The time needed to achieve a
state that the angular momentum profile is smooth throughout the
flow is approximately the viscous timescale at the outer boundary.
Therefore, for an accretion system which can persist for a time that
is longer than this timescale, the radial profile of the angular
momentum will eventually become smooth throughout the accretion
flow. On the other hand, we would like to emphasize that for
transient accretion systems, such as the tidal disruption of a star
by a black hole, the radial profile of the angular momentum should
be smooth only inside a finite radius $R_0$, at which the viscous
timescale equals the life time of the accretion event. Within this
radius the specific angular momentum of the accretion flow is close
to Keplerian or sub-Keplerian while beyond $R_0$ the angular
momentum should roughly be constant.

In this picture, the angular momentum is significantly smaller than
the Keplerian value, which implies that the gravitational force is
larger than the centrifugal force.  For a hot accretion flow, such a
difference of force is balanced by the gradient of gas pressure.
While for a cold accretion flow, the gradient of gas pressure may
not be large enough to play such a role. In this case, perhaps it is
the inertia force together with the centrifugal force that balances
the gravitational force. Our simulation presented in this paper only
holds for hot accretion flows. The simulation of cold accretion disk
is technically difficult for us thus needs to be checked  in the
future.

\section*{ACKNOWLEDGMENTS}

We thank the anonymous referee for the constructive comments. We
thank C. K. Chan and X. H. Yang for useful discussions. This work
was supported in part by the Natural Science Foundation of China
(grants 11103061, 11133005, 11121062, and 11103059), the National
Basic Research Program of China (973 Program, grant 2014CB845800),
and the Strategic Priority Research Program ¡°The Emergence of
Cosmological Structures¡± of the Chinese Academy of Sciences (grant
XDB09000000). The simulations were carried out at the Super
Computing Platform of Shanghai Astronomical Observatory.

\end{document}